\documentclass[a4paper, amsfonts, amssymb, amsmath, reprint, showkeys, nofootinbib, twoside,aps,prl]{revtex4-2}
\usepackage[english]{babel}
\usepackage[utf8]{inputenc}
\usepackage[colorinlistoftodos, color=green!40, prependcaption]{todonotes}
\usepackage{xcolor}
\usepackage{xspace}
\usepackage{lineno}
\usepackage{amssymb}
\usepackage{url}
\usepackage{ulem}
\usepackage{amsthm}
\usepackage{mathtools}
\usepackage{physics}
\usepackage{xcolor}
\usepackage{graphicx}
\usepackage[left=23mm,right=13mm,top=35mm,columnsep=15pt]{geometry} 
\usepackage{adjustbox}
\usepackage{placeins}
\usepackage[T1]{fontenc}
\usepackage{lipsum}
\usepackage{csquotes}
\usepackage[Symbolsmallscale]{upgreek}
\newcommand{\pp}           {pp\xspace}

\newcommand{\pP}{\ensuremath{\mbox{p--p}}\,}

\newcommand{\kstar}        {\ensuremath{k^*}\xspace}
\newcommand{\rstar}     {\ensuremath{r^{*}}\xspace}

\newcommand{\nineH}        {$\sqrt{s}~=~0.9$~Te\kern-.1emV\xspace}
\newcommand{\seven}        {$\sqrt{s}~=~7$~Te\kern-.1emV\xspace}
\newcommand{\onethree}        {$\sqrt{s}~=~13$~Te\kern-.1emV\xspace}
\newcommand{\twoH}         {$\sqrt{s}~=~0.2$~Te\kern-.1emV\xspace}
\newcommand{\twosevensix}  {$\sqrt{s}~=~2.76$~Te\kern-.1emV\xspace}
\newcommand{\five}         {$\sqrt{s}~=~5.02$~Te\kern-.1emV\xspace}
\newcommand{\twosevensixnn}{$\sqrt{s_{\mathrm{NN}}}~=~2.76$~Te\kern-.1emV\xspace}
\newcommand{\fivenn}       {$\sqrt{s_{\mathrm{NN}}}~=~5.02$~Te\kern-.1emV\xspace}

\newcommand{\GeVc}         {Ge\kern-.1emV/$c$\xspace}
\newcommand{\MeVc}         {Me\kern-.1emV/$c$\xspace}
\newcommand{\GeVmass}      {Ge\kern-.1emV/$c^2$\xspace}
\newcommand{\MeVmass}      {Me\kern-.1emV/$c^2$\xspace}

\newcommand{\ks}     {\ensuremath{k^{*}}\xspace}

\newcommand{\SPi}         {\ensuremath{\uppi\Sigma}\xspace}
\newcommand{\kbarN}         {\ensuremath{\rm\overline{K}N}\xspace}
\newcommand{\kMinProt}         {\ensuremath{\rm K^{-}p}\xspace}

\newcommand{\LAK}{${\mathrm{\overline{K}}} \Lambda$\xspace} 
\newcommand{\XiPi}{${\uppi} \Xi$\xspace} 
\newcommand{\SIK}{${\mathrm{\overline{K}}} \Sigma$\xspace} 
\newcommand{\XiEta}{${\eta} \Xi$\xspace}

\newcommand{\LKMin}{\ensuremath{\rm K^{-} \Lambda}\xspace}

\newcommand{\PiMinXiZ}{\ensuremath{\rm \uppi^- \Xi^0 }\xspace}
\newcommand{\PiZXiMin}{\ensuremath{\rm \uppi^0 \Xi^- }\xspace}

\newcommand{\KMinSigZ}{\ensuremath{\rm K^- \Sigma^0}\xspace}
\newcommand{\KZbarSigMin}{\ensuremath{ \overline{\rm K}^0 \Sigma^-}\xspace}
\newcommand{\EtaXiMin}{\ensuremath{\rm \eta \Xi^- }\xspace}

\newcommand{\XRes}          {\ensuremath{\Xi\mathrm{(1620)}}\xspace}
\newcommand{\XResNovanta}          {\ensuremath{\Xi\mathrm{(1690)}}\xspace}
\newcommand{\XResVenti}          {\ensuremath{\Xi\mathrm{(1820)}}\xspace}

\usepackage[pdftex, pdftitle={Article}, pdfauthor={Author}]{hyperref} 
\begin{document}
\title{Constraining the low-energy $S=-2$ meson-baryon interaction\\ with two-particle correlations}

\author{V. Mantovani Sarti$^{a,*}$, A. Feijoo$^{b,\dagger}$, I. Vida\~na$^{c,\ddagger}$, A. Ramos$^{d,\S}$, F. Giacosa$^{e,f}$, T. Hyodo$^{g,h}$, Y. Kamiya$^{i,h}$}
    \affiliation{$^{a}$ Physik Department E62, Technische Universit\"at M\"unchen, Garching, Germany, EU}
        \email{valentina.mantovani-sarti@tum.de}
    \affiliation{$^{b}$Instituto de F\'{i}sica Corpuscular, Centro Mixto Universidad de Valencia-CSIC, Institutos de Investigaci\'{o}n de Paterna, Aptdo. 22085, E-46071 Valencia, Spain}
        \email{edfeijoo@ific.uv.es}
    \affiliation{$^{c}$
    Istituto Nazionale di Fisica Nucleare, Sezione di Catania, Dipartimento di Fisica ``Ettore Majorana'', Universit\`a di Catania, Via Santa Sofia 64, I-95123 Catania, Italy}
    \email{isaac.vidana@ct.infn.it}
\affiliation{$^{d}$Departament de F\'{i}sica Qu\`antica i Astrof\'{i}sica and Institut de Ci\`encies del Cosmos (ICCUB), Facultat de F\'{i}sica, Universitat de Barcelona, Barcelona, Spain}
\email{ramos@fqa.ub.edu}
    \affiliation{$^{e}$
   Institute of Physics, Jan Kochanowski University, ul. Uniwersytecka 7, 25-406 Kielce, Poland}   
    \affiliation{$^{f}$
   Institute of Theoretical Physics, Goethe University, Max-von-Laue-Str.1, 60438, Frankfurt am Main, Germany}    
    \affiliation{$^{g}$
    Department of Physics, Tokyo Metropolitan University, Hachioji 192-0397, Japan
    }
    \affiliation{$^{h}$
    RIKEN Interdisciplinary Theoretical and Mathematical Science Program (iTHEMS), Wako 351-0198, Japan
    }
    \affiliation{$^{i}$Helmholtz-Institut f\"ur Strahlen- und
	Kernphysik and Bethe Center for Theoretical Physics,\\
	Universit\"at Bonn, D-53115 Bonn, Germany}


\begin{abstract}
The two-particle correlation technique applied to \LKMin pairs in \pp collisions at LHC recently provided the most precise data on the strangeness $S=-2$ meson-baryon interaction. In this letter, we use for the first time femtoscopic data to constrain the parameters of a low-energy effective QCD Lagrangian. The tuned model delivers new insights on the molecular nature of the \XRes and \XResNovanta states. This procedure opens the possibility to determine higher order corrections, directly constraining QCD effective models particularly in the multi-strange and charm sectors.

\end{abstract}


\maketitle

\textit{Introduction:} The dynamics of the strong interaction between strange hadrons at low and intermediate energies is still a rather uncharted territory, both experimentally and theoretically.
Namely, in this energy regime, a quantitative description of hadronic interactions in terms of the elementary quark and gluon degrees of freedom is hindered by the mathematical problems associated to the non-perturbative character of Quantum-Chromo-Dynamics (QCD). Effective Lagrangians, employing hadronic degrees of freedom and including the fundamental symmetries of QCD as well as spontaneous and anomalous breaking patterns, have been widely used to circumvent this difficulty since the pioneering work of Weinberg \cite{Weinberg:1978kz,Weinberg90,Weinberg91}.\\
Amongst these effective approaches, Chiral Perturbation theory ($\chi$PT) has proved to be an extremely powerful tool to compute in a systematic way many low-energy observables (e.g. cross-sections) and to provide insights on the underlying hadronic interactions. The $\chi$PT framework allows us to investigate higher-order terms in the chiral expansion of the interaction, which can significantly improve the understanding of the underlying QCD dynamics in the system at hand. 
Each term in the Lagrangian is preceded by the so-called Low Energy Constants (LECs), parameters which are not fixed by the underlying theory and hence must be determined by a fit to the available experimental data.\\ 
In this work we focus on the $S=-2$ meson-baryon interaction, dominated by the \XiPi--\LAK--\SIK--\XiEta coupled-channel dynamics. 
Similarly to the $\Lambda(1405)$, a molecular \SPi--\kbarN state arising from the interplay between these two coupled-channels~\cite{1405_reviewHyodo,Meissner:2020khl,1405_reviewMai,Hyodo:2020czb}, the \XRes and \XResNovanta resonances might also be dynamically generated within the $S=-2$ meson-baryon interaction, thereby acquiring a molecular structure.
In spite of the large theoretical effort devoted in the last two decades~\cite{Ramos02,Garcia-Recio1,Garcia-Recio2,Sekihara:2015qqa,Khemchandani:2016ftn,Nishibuchi:2023acl} to the understanding 
of the properties of the \XRes and \XResNovanta states, a clear picture on their nature is still missing. Recently, the authors of Ref.~\cite{Feijoo:2023wua} delivered the first unitarized effective meson-baryon chiral Lagrangian that includes contributions beyond the leading contact-term interaction in the $S=-2$ sector.
The inclusion of higher orders in the chiral expansion reduced the disagreement
between model predictions and the available experimental data 
on the two $\Xi^{\ast}$ states, but it also introduced more LECs, namely more unknown parameters to be constrained from data. Unfortunately, current experimental information, which includes only evidence of the \XResNovanta resonance decaying into \LKMin\cite{LHCb,Belle:2001hyr,BESIII} and the first observation of the neutral \XRes decaying into $\uppi^+\Xi^-$~\cite{BELLE}, is not sufficient to constrain the next-to-leading order (NLO) LECs. This limitation drove the authors of Ref.~\cite{Feijoo:2023wua} to invoke $SU(3)$ flavour symmetry
to adopt the same LECs of the meson-baryon interaction models in the $S=-1$ sector, which benefit from a much larger data sample~\cite{Humphrey:1962zz,Watson:1963zz,Mast:1975pv,Nowak:1978au,Ciborowski:1982et,Sakitt:1965kh,Bazzi11,Bazzi12,Lambda1405_1,Lambda1405_2,Lambda1405_3,Lambda1405_4,Lambda1405_CLAS}. Clearly, an improvement in the theoretical developments on the $S=-2$ meson-baryon interaction requires new and more precise experimental data.\\
Such experimental input became available with the recent ALICE measurement of the \LKMin correlation function (CF) in \pp collisions at \onethree, which delivered the most precise data on this interaction and provided the first experimental observation of the \XRes decaying into \LKMin pairs~\cite{LKALICEpp}.
In the present study, these femtoscopic data is used for the first time to fit the parameters of the state--of--the--art $\chi$PT effective Lagrangian at NLO~\cite{Feijoo:2023wua}. The resulting unitarized $\chi$PT (U$\chi$PT) amplitudes 
are employed to investigate the position and couplings of the poles related to the \XRes and \XResNovanta states to the different channels, leading to a complete new distribution of the molecular composition of these two resonances.\\ 
The results presented in this work make use of a novel method to constrain the low-energy QCD effective models for interactions involving multi-strange and charm hadrons, which cannot be accessed via traditional scattering experiments. For these interactions, correlation measurements at LHC already provided the experimental access to a large amount of two-body and three-body interactions~\cite{ALICE:Run1,ALICE:pXi,ALICE:pSig0,ALICE:LL,ALICE:pOmega,ALICE:pKpp,ALICE:pKCoupled,ALICE:pLCoupled,ALICE:BBar,ALICE:LXi} In the future, thanks to the even larger statistics expected at LHC~\cite{ALICERun3Run4}, and with brand new dedicated experiments~\cite{ALICE3}, femtoscopic measurements will be the only data at our disposal able to directly constrain the two-body scattering amplitude, hence the methodology described in this Letter provides a tool to guide future comparisons with correlation data.\\ 
\newline
\textit{Formalism:}
In the case of a multi-channel system, such as the one we are considering in this work, the corresponding two-particle CF of a given observed channel $i$ (e.g. \LKMin) reads~\cite{LisaPrattReview,Haidenbauer:CC,Kamiya:CCKp}
\begin{align}\label{eq:corrfun_CC}
    C_i(\ks) = \sum_{j} \omega_{j} ^{\rm prod.} \int d^3 \rstar S_{j} (\rstar) |\psi_{ji} (\ks,r^*)|^2.
\end{align}
Here \kstar and \rstar represent, respectively, the relative momentum and distance between the two particles, measured in the pair rest frame. The sum runs over the elastic ($j=i=\LKMin$) and the inelastic channels ($j = \PiMinXiZ, \PiZXiMin, \KMinSigZ, \KZbarSigMin, \EtaXiMin$).\\
The contributions of the inelastic channels are scaled by the production weights $\omega_{j} ^{\rm prod.}$, which take into account how many $j$ pairs, produced as initial states, can convert to the measured $i$ final state. The emitting source $S_{j}(\rstar)$ describes the probability of emitting the $j$ pair at a relative distance \rstar and, particularly in \pP femtoscopic measurements, might be different in each channel due to the feed-down from strongly decaying resonances, specific for each pair~\cite{ALICE:Source,ALICE:pKCoupled,CECA}.
Finally, the last ingredient is the relative wave function $\psi_{ji} (\ks,r^*)$, embedding the strong interaction arising from the coupled-channel dynamics in the system. Following the formalism in~\cite{Haidenbauer:CC}, the wave functions can be obtained from the scattering amplitude.\\
The starting point from which we derive the scattering amplitude within the U$\chi$PT is the chiral effective Lagrangian up to NLO $\mathcal{L}_{\phi B}^{eff}=\mathcal{L}_{\phi B}^{(1)}+\mathcal{L}_{\phi B}^{(2)}$, with
\begin{eqnarray} 
\mathcal{L}_{\phi B}^{(1)} & = & i \langle \bar{B} \gamma_{\mu} [D^{\mu},B] \rangle
                            - M_0 \langle \bar{B}B \rangle  
                           - \frac{1}{2} D \langle \bar{B} \gamma_{\mu} 
                             \gamma_5 \{u^{\mu},B\} \rangle \nonumber \\
                  & &      - \frac{1}{2} F \langle \bar{B} \gamma_{\mu} 
                               \gamma_5 [u^{\mu},B] \rangle \ ,
\label{LagrphiB1} \\
  \mathcal{L}_{\phi B}^{(2)}& = & b_D \langle \bar{B} \{\chi_+,B\} \rangle
                             + b_F \langle \bar{B} [\chi_+,B] \rangle
                             + b_0 \langle \bar{B} B \rangle \langle \chi_+ \rangle  \nonumber \\ 
                     &  & + d_1 \langle \bar{B} \{u_{\mu},[u^{\mu},B]\} \rangle 
                            + d_2 \langle \bar{B} [u_{\mu},[u^{\mu},B]] \rangle     \nonumber \\
                    &  &  + d_3 \langle \bar{B} u_{\mu} \rangle \langle u^{\mu} B \rangle
                            + d_4 \langle \bar{B} B \rangle \langle u^{\mu} u_{\mu} \rangle \ .
\label{LagrphiB2}
\end{eqnarray}
The contact term, corresponding to the Weinberg-Tomozawa (WT) contribution, and the direct and crossed Born terms are included in $\mathcal{L}_{\phi B}^{(1)}$ whereas the tree-level NLO contributions are fully extracted from $\mathcal{L}_{\phi B}^{(2)}$.
In these equations, $B$ is the octet baryon matrix, while the matrix of the pseudoscalar mesons $\phi$  is implicitly contained in 
$u_\mu = i u^\dagger \partial_\mu U u^\dagger$, where $U(\phi) = u^2(\phi) = \exp{\left( \sqrt{2} {\rm i} \phi/f \right)} $ with $f$ being the effective meson decay constant. 
The covariant derivative is given by  $[D_\mu, B] = \partial_\mu B + [ \Gamma_\mu, B]$, with $\Gamma_\mu =  [ u^\dagger,  \partial_\mu u] /2$, while $\chi_+ = 2 B_0 (u^\dagger \mathcal{M} u^\dagger + u \mathcal{M})$, with $\mathcal{M} = {\rm diag}(m_u, m_d, m_s)$ and $B_0 = - \langle 0 |\bar{q} q | 0 \rangle / f^2$, is the explicit chiral symmetry breaking term. The values of the axial vector constants are taken as $D=0.8$ and $F=0.46$ and $M_0$ is the baryon octet mass in the chiral limit. The NLO Lagrangian depends on a few LECs, namely $b_D$, $b_F$, $b_0$ and $d_i$ $(i=1,\dots,4)$, to be determined here from the fit to the measured \LKMin correlation.\\
The total interaction kernel up to NLO, derived from Eqs.~(\ref{LagrphiB1}) and (\ref{LagrphiB2}), reads $V_{ij}=V^{\scriptscriptstyle WT}_{ij}+V^{D}_{ij}+V^{C}_{ij}+V^{\scriptscriptstyle NLO}_{ij}$, where the elements of the  interaction matrix $\hat{V}_{ij}$ couple all possible meson-baryon channels, see Refs.~\cite{Feijoo:2021zau,Ramos:2016odk} for details.
The interaction kernel in the present ($S=-2$, $Q=-1$) sector is derived from the ($S=-2$, $Q=0$) one~\cite{Feijoo:2023wua} by employing basic isospin arguments.\\
The final step is to connect the interaction kernel to the scattering amplitude $T_{ij}$, required to evaluate the wave functions and calculate the CF. The U$\chi$PT method, adopted here due to the presence of the $\Xi^*$ resonances, solves the Bethe--Salpeter equations through an on-shell factorization, leaving a simple system of algebraic equations expressed in matrix form as
\begin{equation}
T_{ij} ={(1-V_{il}G_l)}^{-1}V_{lj} ,
 \label{T_algebraic}
\end{equation}
being $G_l$ the meson-baryon loop function whose logarithmic divergence is handled by dimensional regularization
\begin{eqnarray}
 G_l & = &\frac{2M_l}{(4\pi)^2} \Bigg \lbrace a_l(\mu)+\ln\frac{M_l^2}{\mu^2}+\frac{m_l^2-M_l^2+s}{2s}\ln\frac{m_l^2}{M_l^2}  \nonumber \\ 
 &  +   &  \frac{q_{\rm cm}}{\sqrt{s}}\ln\left[\frac{(s+2\sqrt{s}q_{\rm cm})^2-(M_l^2-m_l^2)^2}{(s-2\sqrt{s}q_{\rm cm})^2-(M_l^2-m_l^2)^2}\right]\Bigg \rbrace.  
 \label{dim_reg}    
\end{eqnarray}
The former expression comes in terms of the baryon ($M_l$) and meson ($m_l$) masses for the $l$-channel as well as the subtraction constants (SCs) $a_l$, replacing the divergence for a given dimensional regularization scale $\mu$, taken to be $1$~GeV. Despite a natural size can be established for them, the lack of knowledge about the SCs requires their inclusion in the fitting procedure. The number of independent SCs is four following isospin symmetry arguments. 
Summarizing, the U$\chi$PT with WT+Born+NLO terms in this sector leaves a scattering amplitude that depends on $13$ parameters never determined before. Hence, the \LKMin CF offers an unprecedented opportunity to constrain a theoretical model that can be employed to make novel predictions in a quite unknown sector.\\
The procedure to fit the LECs and SCs of this model, referred from now on as the Valencia--Barcelona--Catania (VBC) model, is described in the following section.\\

\newline
\textit{Fitting procedure:} Following the approach in the experimental \LKMin analysis~\cite{LKALICEpp}, the function we use to fit the data reads
\begin{align}\label{eq:totcorrelation}
    C(\kstar) = & N_D \times C_\mathrm{model}(\kstar) \times C_\mathrm{background} (\kstar).
\end{align}
The term $C_\mathrm{model}(\kstar)=1 + \lambda_{\rm gen} \times (C_{\rm gen}(\kstar) - 1)+ \sum_{\rm res} \lambda_{\rm res} \times (C_{\rm res}(\kstar) - 1)$ includes the genuine \LKMin correlation, defined via Eq.~(\ref{eq:corrfun_CC}) and obtained within the VBC model, as well as the residual contributions.
The correlations are weighted with the same $\lambda$ parameters used in~\cite{LKALICEpp} and the interaction amongst the pairs composing the residual correlations is modeled using the same assumptions adopted in~\cite{LKALICEpp}.\\
A prior knowledge of the source function is needed to evaluate both the genuine and residual contributions~\cite{femtoreview,LisaPrattReview}. 
In particular, to obtain $C_{\rm gen}(\kstar)$, we determined the specific emitting sources for each elastic and inelastic channel, following the same approach used in the treatment of the coupled-channel contributions of the \kMinProt correlation in~\cite{ALICE:pKCoupled}. For the elastic \LKMin part, we assume the same double Gaussian source  parametrization employed in~\cite{LKALICEpp}. The same source distribution is assumed also for the residual correlations.\\
As shown in~\cite{ALICE:pKCoupled}, depending on the pairs entering the inelastic contributions, the corresponding source profiles can significantly deviate from the elastic one. Such an effect is particularly relevant when pions are involved, as in the case at hand here in which the channels \PiMinXiZ and \PiZXiMin are present. We perform a detailed study on the source profiles of the inelastic channels by adopting the data-driven resonance source model (RSM) in~\cite{ALICE:Source}, used as well in the \LKMin correlation measurement~\cite{LKALICEpp} and in several femtoscopic analyses~\cite{ALICE:pOmega,ALICE:LXi,ALICE:BBar,ALICE:pLCoupled,ALICE:pXi,ALICE:pD,ALICE:pKCoupled}. 
Following 
~\cite{ALICE:pKCoupled}, we build, for each inelastic channel, a total source having a Gaussian core with radius $r_\mathrm{core}=1.11\pm0.04$~\cite{LKALICEpp}, common to all channels, and a non-gaussian contribution from the feeding of strongly decaying resonances. As done for the elastic \LKMin channel, the final inelastic sources are modeled with a double Gaussian parametrization. 
Compatible parameters are found between the \KMinSigZ and \KZbarSigMin effective sources and the \LKMin one, while larger radii are obtained for the $\uppi\Xi$ channels due to the long-lived resonances ($c\tau \gtrsim 5$ fm) feeding to the pions. The remaining $S_{\EtaXiMin}(\rstar)$ distribution is localized at slightly smaller \rstar since no significant strong feed-down to the $\Xi$ baryon is present.
We assign the same relative uncertainties ($\approx4\%$) on the inelastic source´s parameters of the reported one on $r_{\mathrm{core}}$ in~\cite{LKALICEpp}.\\
The last quantities to be evaluated are the production weights $\omega_{j} ^{\rm prod.}$. We employed the same data-driven method used for the \kMinProt correlation analysis~\cite{ALICE:pKCoupled}, measured in the same high-multiplicity dataset we are considering. 
The values obtained for each inelastic channel (normalized to the \LKMin one, e.g. $\omega_{\LKMin} ^{\rm prod.} = 1.$) are $\omega_{\PiMinXiZ} ^{\rm prod.} = 1.53$, $\omega_{\PiZXiMin} ^{\rm prod.} = 1.58$, $\omega_{\KMinSigZ} ^{\rm prod.} = 0.68$, $\omega_{\KZbarSigMin} ^{\rm prod.} = 0.63$ and $\omega_{\EtaXiMin} ^{\rm prod.} = 0.18$. 
We associate errors on the productions weights of the order of maximum $\approx10\%$ by propagating the available uncertainties on the parameters used to estimate the inelastic pairs yields~\cite{TFMultiplicity} and the \kstar kinematics, as done in~\cite{ALICE:pKCoupled}. Additional details on the source and $\omega_{j} ^{\rm prod.}$ determination can be found in the Supplemental Material.\\
For the comparison with the measured \LKMin correlation in~\cite{LKALICEpp}, a residual background, given by the last term in Eq.~(\ref{eq:totcorrelation}), must be taken into account. 
The profile of $C_\mathrm{background} (\kstar)$, publicly available at~\cite{hepdata.143518}, is composed of a polynomial function plus the presence of several resonances at large \kstar, amongst which the \XResNovanta. In~\cite{LKALICEpp}, the latter has been modeled with a Breit-Wigner distribution and the corresponding mass and widths extracted from the fit were found compatible with the PDG values~\cite{PDG}. In our work, the \XResNovanta is dynamically generated within the VBC model entering the $C_\mathrm{model} (\kstar)$ term, hence the modeling of this resonance in the background should not be included. To do so, we perform a fit to the available $C_\mathrm{background} (\kstar)$, using Eq.~(3) in~\cite{LKALICEpp}, and we set to zero the Breit-Wigner term for the \XResNovanta. The final background correlation entering in our fit contains only as resonances the $\Omega$ at $\kstar \approx 210$ \MeVc and the \XResVenti at $\kstar \approx 400$ \MeVc.\\
The fit of $C_\mathrm{tot} (\kstar)$ to the \LKMin correlation data is performed, as in~\cite{LKALICEpp}, in the range $0 \leq \kstar \leq 500$ \MeVc, leaving as free parameters the normalization constant $N_D$, the LECs and SCs of the VBC model.
The fitting procedure we adopt is based on the bootstrap technique used as in~\cite{LKALICEpp}. A total of 1000 samplings is performed in which we also, at each iteration, vary randomly the values of the source parameters and production weights within the quoted uncertainties.\\

\begin{figure}[t!]
\centering
	\includegraphics[width=0.49\textwidth]{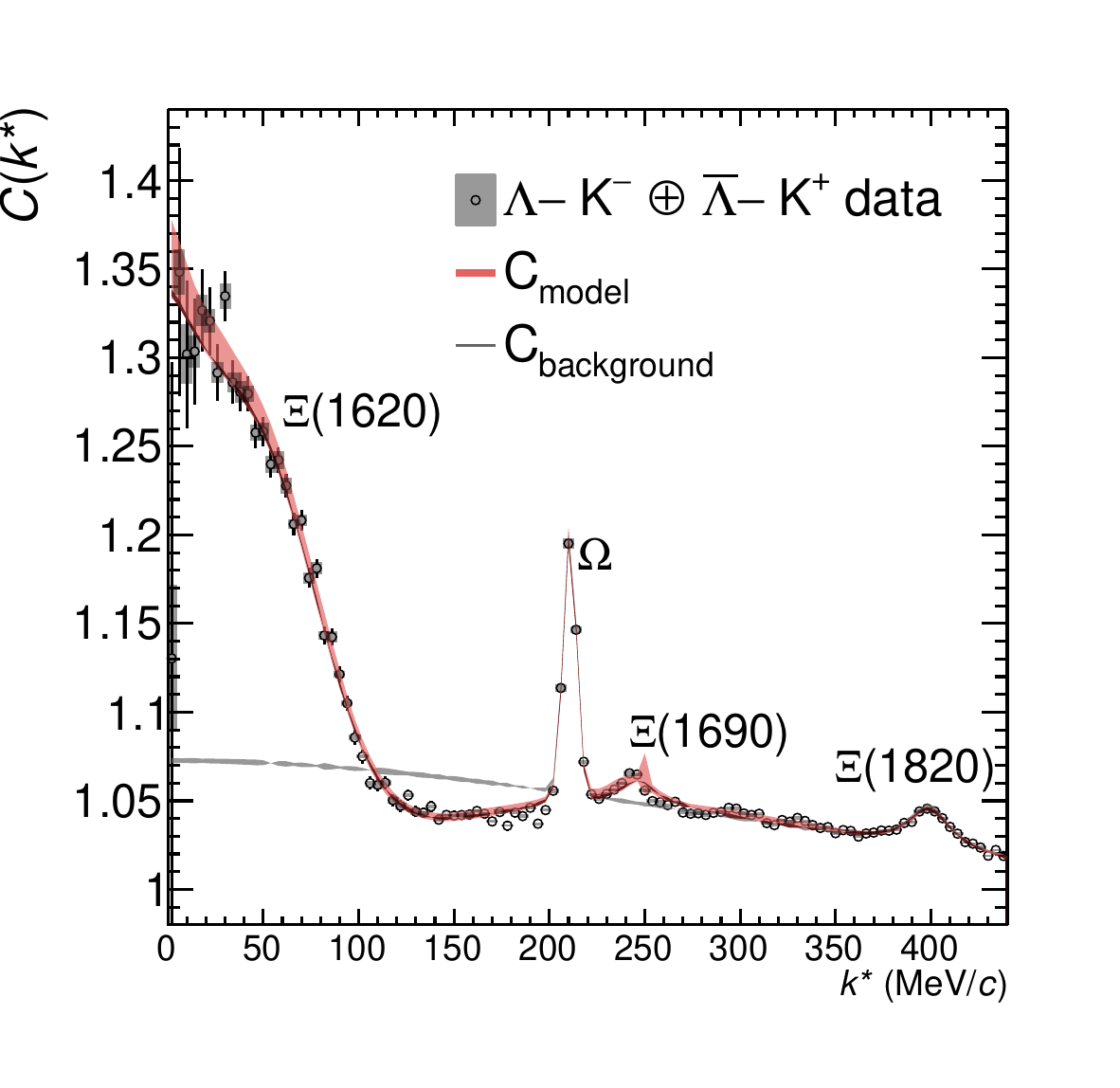}
\caption{Upper panel: Experimental \LKMin correlation data with systematic (gray boxes) and statistical (vertical lines) uncertainties~\cite{LKALICEpp}. The red band is the total fit obtained with the VBC model. The darker shade is due to the statistical uncertainty on the data, while the light shade corresponds to the total error $\sigma_{\rm tot} = \sqrt{\sigma_{\rm stat}^2 + \sigma_{\rm syst}^2}$. The gray band represents the background. Lower panel: deviation between data and model in terms of numbers of standard deviations.}
\label{fig:ExpCF}
\end{figure}

\newline
\textit{Results:} In Fig.~\ref{fig:ExpCF} we present in the upper panel the results of the fit of $C(\kstar)$ (red bands) to the measured \LKMin correlation. In the lower panel we estimate for each \kstar interval the agreement between the data and the model (normalized to the statistical error of the data), expressed in terms of numbers of standard deviation (n$\sigma$). In the considered \kstar range the average n$\sigma$ is around $1.3$, confirming the agreement between the femtoscopic data and the tuned VBC model.\\
The extracted parameters of the VBC model are shown in Tab.~\ref{tab:param}. 
Values of the SCs of $\sim-2$ are of `natural size'~\cite{Oller:2000fj}. The lowest-order LEC $f$ tends towards its smallest allowed value, $f_\pi$. The NLO LECs are, in general, comparable in size with those determined for the $S=-1$ interaction~\cite{Borasoy:2005ie,Guo:2012vv,Feijoo:2018den}, with the exception of $b_0$, whose size turns out to be roughly one order of magnitude larger. 
As this parameter appears in all the diagonal elements of the NLO $D_{ij}$ coefficients (see Table~1 in~\cite{Feijoo:2023wua} and Eq.~(10) in~\cite{Ramos:2016odk}), it is responsible for the generation of moderately attractive interactions in the \LKMin and \EtaXiMin channels that are otherwise null at the WT level. This provides the $S=-2$ model at NLO with a richer coupled-channel structure than its LO counterpart, allowing it to reproduce the CF in a wide momentum range.
\begin{table}[ht]
\centering
\caption{Extracted fit parameters with LECs and subtraction constants of the VBC $S=-2$ meson-baryon interaction at NLO and normalization constant $N_D$. The bootstrap method~\cite{PresTeukVettFlan92,Efron:1986hys,Albaladejo:2016hae} was employed to determine the errors of the parameters.}
\begin{tabular}{l | r  }
\hline 
\hline 
$a_{ \Xi\pi}$  &  $-2.96 \pm  0.11$ \\
$a_{\Lambda\overline{\text{K}}}$  &  $  -1.87   \pm   0.10 $  \\
 $a_{\Sigma\overline{\text{K}}}$  & $ -1.32   \pm    0.02 $ \\
 $a_{\Xi \eta}$ & $-2.42 \pm     0.03 $   \\[1mm]
\hline
\\[-4mm]  
 $f/f_\pi$ & $1.000 \pm     0.001 $   \\[1mm]
\hline
\\[-4mm]  
$b_0$ [GeV$^{-1}]$ &  $ -2.997  \pm  0.002$  \\
$b_D$ [GeV$^{-1}]$ &  $ 1.20 \pm   0.09$ \\
$b_F$ [GeV$^{-1}]$ &  $ -0.30\pm   0.12$ \\
$d_1$ [GeV$^{-1}]$ &  $ -0.69 \pm  0.18$  \\
$d_2$ [GeV$^{-1}]$ &  $ -0.21 \pm  0.06$ \\
$d_3$ [GeV$^{-1}]$ &  $ 0.08\pm   0.20$ \\
$d_4$ [GeV$^{-1}]$ &  $ -0.39 \pm   0.05$  \\                 
\hline
\\[-4mm]  
$N_D$  &  $ 1.0015  \pm   0.0004$ \\[1mm]
\hline
\hline
\hline
\end{tabular}
\label{tab:param}
\end{table}
Indeed, the tuned VBC model describes the data very well in the region of $\kstar \leq 200$ \MeVc, where the presence of the \XRes, dynamically generated as a meson-baryon quasi-bound state in the model, is dominant. A reasonable description is also found for the peak around $\kstar \approx 250$ \MeVc associated to the \XResNovanta state, which is also generated dynamically by the VBC model. 

In Table~\ref{tab:poles} we present the pole properties for the \XRes and \XResNovanta states obtained in the VBC model. Both poles are found in the physically relevant Riemann sheet and the corresponding masses and widths are compatible with the current experimental data~\cite{BELLE,LHCb}, reported in the last two rows. Such an agreement confirms, as demonstrated in~\cite{Feijoo:2023wua} and in contrast to all previous studies~\cite{Ramos02,Garcia-Recio1,Garcia-Recio2,Sekihara:2015qqa,Khemchandani:2016ftn,Nishibuchi:2023acl}, that the inclusion of Born and the NLO contributions is crucial to dynamically generate both $\Xi^\ast$ simultaneously.
\begin{table}[t!]
\centering
\caption{Poles, couplings and compositeness of the resonances generated by the VBC $S=-2$ meson-baryon interaction at NLO. The number between brackets in the first column denotes the channel threshold energy in MeV.}
\begin{tabular}{l|cc|cc}
\multicolumn{5}{c}{}\\[-2.5mm]
\hline
\hline \multicolumn{5}{c}{}\\[-3.5mm]
 \multicolumn{1}{c}{mass $M$:}  &  \multicolumn{2}{c}{$1612.68$ MeV}  &  \multicolumn{2}{c}{$1670.28$ MeV}   \\
 \multicolumn{1}{c}{width $\Gamma$:}   &       \multicolumn{2}{c}{$24.57$ MeV} &      \multicolumn{2}{c}{$7.44$ MeV} \\ 
 \multicolumn{1}{c}{Riemann sheet:}     &      \multicolumn{2}{c}{$(---+++)$} & \multicolumn{2}{c}{$(---+++)$} \\
 \multicolumn{5}{c}{}\\[-2mm]
  \multicolumn{1}{c}{}     &   $| g_i|$   &  \multicolumn{1}{c}{$\left|g_i^2 dG/dE\right| $}   &  \multicolumn{1}{c}{$| g_i|$}   & $\left|g_i^2 dG/dE\right| $  \\[1mm]
                              \hline
$ \PiMinXiZ (1454)$    &   0.51 & $   0.014$   &   0.22 & $   0.002$    \\
$ \PiZXiMin (1456)$    &   0.36 & $   0.007$    &   0.39 & $   0.007$    \\
$\LKMin (1609)$ &  0.94  & $   0.162    $   &  0.07  & $   0.000   $    \\
$\KMinSigZ (1686)$   &    0.17   & $   0.002$    &    2.20   & $   0.761$       \\
$\KZbarSigMin (1695)$   &    0.21   & $   0.003$ &    1.37  & $   0.230$       \\
$\EtaXiMin (1868)$   &    5.86 & $   0.937$   &    0.05 & $   0.000$   \\
\hline
 \multicolumn{1}{c}{Experimental $\Xi^*$:}     &      \multicolumn{2}{c}{\XRes~\cite{BELLE}} & \multicolumn{2}{c}{\XResNovanta~\cite{PDG}} \\
 \multicolumn{1}{c}{mass $M$:}  &  \multicolumn{2}{c}{$1610.4\pm6.0^{+5.9}_{-3.5}$ MeV}  &  \multicolumn{2}{c}{$1690\pm10$ MeV}   \\
 \multicolumn{1}{c}{width $\Gamma$:}   &       \multicolumn{2}{c}{$59.9\pm4.8^{+2.8}_{-3.0}$ MeV} &      \multicolumn{2}{c}{$20\pm15$ MeV} \\ 
\hline
 \multicolumn{5}{c}{}\\[-4.2mm]
\hline
\end{tabular}
\label{tab:poles}
\end{table}
A novel aspect of the present study comes when inspecting the couplings $g_i$ of the different channels to the \XRes pole. The strong coupling to the highest channel, at the expense of reducing sizeably the ones to the \XiPi and \LKMin, reveals a paradigm shift in the compositeness of the \XRes state. All former works interpret such a state as a \XiPi--\LAK molecule with a non-negligible coupling to the \SIK channel. In the present study, the molecule basically consists of a \LKMin--\EtaXiMin mixing, with the latter being the dominant component. A direct consequence of the reduced couplings to the \XiPi and \LKMin open channels is a much narrower \XRes width, in contrast to previous values, which approaches the experimental data. For the \XResNovanta resonance, which appears basically as a \SIK quasi-bound state, we observe that the theoretical energy is located below the expected experimental value and the lowest \SIK threshold. The latter condition reduces the possibility of decaying into \KMinSigZ states, leading to a reduction of the width with respect to~\cite{Feijoo:2023wua}.  

A good quality fit to the CF at low momenta can also be obtained with the simpler WT model at the expense of some SCs being close to zero and thus being rather `unnatural'. The resulting amplitudes present a low energy pole compatible with the lower one of the NLO fit. However, the WT model fails at describing the data around $k^*\sim 250$~MeV/c since no higher energy pole is generated.

Two important observations can be drawn: first,  the information encoded in the CF strongly indicates the existence of a resonance having an energy of around $1620$ MeV and a width of $\lesssim 30$ MeV, as its shape can only be reproduced by an explicit inclusion of the resonance~\cite{LKALICEpp}, or by theoretical models that generate it dynamically with similar characteristics. Secondly, it is necessary to implement the NLO terms of the chiral Lagrangian in order to reproduce the low momenta region of the CF and the \XResNovanta at $k^*\sim 250$ MeV/c.\\
The position of the near-threshold \XRes pole is related to the scattering length $f_{0}$ and effective range $r_{e}$~\cite{Braaten:2004rn,Naidon:2016dpf,Hyodo:2013iga,Nishibuchi:2023acl}.
The tuned VBC model delivers respectively $f_0=(0.23 \pm 0.05)  + \rm{i}\, (0.45 \pm 0.07) $  fm and $r_{\rm eff}=(-6.35 \pm 5.57) + \rm{i}\, (36.22\pm 3.90)  $ fm, in agreement with the results obtained within a more phenomenological approach in~\cite{LKALICEpp}.\\
It is known that a near-threshold resonance above the threshold requires an important contribution from the effective range~\cite{Hyodo:2013iga,Nishibuchi:2023acl}, as it is the case of the \XRes pole with the present model. With the scattering length only, the pole position is estimated as $M-\rm{i}\,\Gamma/2=1739.83 + \rm{i}\,180.53 $ MeV, largely deviating from the values in Table~\ref{tab:poles}. The situation improves by including the effective range correction, $M-\rm{i}\,\Gamma/2=1616.31 + \rm{i}\,1.75 $ MeV. The importance of $r_{e}$, having a larger magnitude than $f_{0}$, is due to the location of the \XRes pole above the \LKMin threshold.
Interestingly, one can also repeat the study of Ref.~\cite{LKALICEpp} by using the same phenomenological Sill  energy line-shape of Ref.~\cite{Giacosa:2021mbz} extended to all the six channels studied in this work. The pole for \XRes in the $(-,-,-,+++)$ Riemann sheet reads $1617.4-\rm{i}\,3.6$ MeV, compatible with (but smaller than) the VBC results.\\ 
\newline
\textit{Conclusions:}
In this letter, we determined for the first time the LECs, up to higher-order corrections, of a state-of-the-art effective Lagrangian from high-precision femtoscopic data.
In particular, we focused on the $S=-2$ meson-baryon sector and use, as experimental constraints, the measured \LKMin correlation by ALICE~\cite{LKALICEpp}.
The description of the data is based on a U$\chi$PT NLO Lagrangian, which accounts for the full $(S=-2,Q=-1)$ meson-baryon coupled channel dynamics and dynamically generates the \XRes and \XResNovanta states~\cite{Feijoo:2023wua}. Effects of the inelastic channels on the calculated \LKMin CF were carefully taken into account with a data-driven estimation on the emitting source parameters and production weights.\\
The VBC model delivers a very good description of the data in the considered \kstar range.
The extracted SCs take their `natural size' values and we observe a large sensitivity of the correlation data to the NLO LECs responsible for the elastic transitions.\\
The fitted parameters were used to study of the \XRes and \XResNovanta poles, whose masses and widths turned out to be compatible with the available experimental measurements. As a novel effect of the femtoscopic constraints and in contrast to previous calculations, one of the molecular states generated (\XRes) mainly consist of a \XiEta quasi-bound state.
\\
The method presented here can be extended to other interactions, involving strange and charm hadrons, which may potentially generate states from coupled channel dynamics, such as the \XRes and \XResNovanta ones in the present study. For these cases the synergy between the theoretical modeling and available femtoscopic data can provide complementary information on the nature of such exotic states.


\section*{Acknowledgements}
This work was supported by
the ORIGINS cluster DFG under Germany’s Excellence
Strategy - EXC2094 - 390783311 and the DFG through
Grant SFB 1258 “Neutrinos and Dark Matter in Astro and Particle Physics”.
V. M. S. is supported by the Deutsche Forschungsgemeinschaft (DFG) through the grant MA $8660/1-1$. This work was also supported by the Spanish Ministerio de Ciencia e Innovaci\'on (MICINN) and European FEDER funds under Contracts  CEX2019-000918-M, PID2020-112777GB-I00, PID2020-118758GB-I00 and by Generalitat Valenciana under contract PROMETEO/2020/023. This project has received funding from the European Union Horizon 2020 research and innovation programme under the program H2020-INFRAIA-2018-1, grant agreement No.\,824093 of the STRONG-2020 project.  A.\,F. is supported through Generalitat Valencia (GVA) Grant APOSTD-2021-112.
This work has been supported in part by the Grants-in-Aid for Scientific Research from JSPS (Grant No. JP22K03637, No. JP19H05150, No. JP18H05402). 
F.G.\ acknowledges support from the \textit{Polish National Science Centre} (NCN) through the \textit{OPUS} project 2019/33/B/ST2/00613.
This work was also supported by the DFG (Project number 196253076 - TRR 110) and the NSFC (Grant No. 11621131001) through the funds provided to the Sino-German CRC 110 ``Symmetries and the Emergence of Structure in QCD" 

\bibliographystyle{utphys}   
\bibliography{bibliography}

\providecommand{\href}[2]{#2}\begingroup\raggedright\begin{thebibliography}{10}

\bibitem{Weinberg:1978kz}
S.~Weinberg, ``{Phenomenological Lagrangians}'',
  \href{http://dx.doi.org/10.1016/0378-4371(79)90223-1}{{\em Physica A}
  {\bfseries 96} no.~1-2, (1979) 327--340}.

\bibitem{Weinberg90}
S.~Weinberg, ``{Nuclear forces from chiral lagrangians}'',
  \href{http://dx.doi.org/10.1016/0370-2693(90)90938-3}{{\em Phys. Lett. B}
  {\bfseries 251} (1990) 288--292}.

\bibitem{Weinberg91}
S.~Weinberg, ``{Effective chiral lagrangians for nucleon-pion interactions and
  nuclear forces}'', \href{http://dx.doi.org/10.1016/0550-3213(91)90231-L}{{\em
  Nucl. Phys. B} {\bfseries 363} (1991) 3--18}.

\bibitem{1405_reviewHyodo}
T.~Hyodo and D.~Jido, ``{The nature of the Lambda(1405) resonance in chiral
  dynamics}'', \href{http://dx.doi.org/10.1016/j.ppnp.2011.07.002}{{\em Prog.
  Part. Nucl. Phys.} {\bfseries 67} (2012) 55--98},
  \href{http://arxiv.org/abs/1104.4474}{{\ttfamily arXiv:1104.4474 [nucl-th]}}.

\bibitem{Meissner:2020khl}
U.-G. Mei\ss{}ner, ``{Two-pole structures in QCD: Facts, not fantasy!}'',
  \href{http://dx.doi.org/10.3390/sym12060981}{{\em Symmetry} {\bfseries 12}
  no.~6, (2020) 981}, \href{http://arxiv.org/abs/2005.06909}{{\ttfamily
  arXiv:2005.06909 [hep-ph]}}.

\bibitem{1405_reviewMai}
M.~Mai, ``{Review of the ${\Lambda }$(1405) A curious case of a strangeness
  resonance}'', \href{http://dx.doi.org/10.1140/epjs/s11734-021-00144-7}{{\em
  Eur. Phys. J. ST} {\bfseries 230} no.~6, (2021) 1593--1607},
  \href{http://arxiv.org/abs/2010.00056}{{\ttfamily arXiv:2010.00056
  [nucl-th]}}.

\bibitem{Hyodo:2020czb}
T.~Hyodo and M.~Niiyama, ``{QCD and the strange baryon spectrum}'',
  \href{http://dx.doi.org/10.1016/j.ppnp.2021.103868}{{\em Prog. Part. Nucl.
  Phys.} {\bfseries 120} (2021) 103868},
  \href{http://arxiv.org/abs/2010.07592}{{\ttfamily arXiv:2010.07592
  [hep-ph]}}.

\bibitem{Ramos02}
A.~Ramos, E.~Oset, and C.~Bennhold, ``{{On the spin, parity and nature of the
  $\Xi$(1620) resonance }}'',
  \href{http://dx.doi.org/10.1103/PhysRevLett.89.252001}{{\em Phys. Rev. Lett.}
  {\bfseries 89} (2002) 252001}.

\bibitem{Garcia-Recio1}
C.~Garcia-Recio, M.~F.~M. Lutz, and J.~Nieves, ``{Quark mass dependence of s
  wave baryon resonances}'',
  \href{http://dx.doi.org/10.1016/j.physletb.2003.11.073}{{\em Phys. Lett. B}
  {\bfseries 582} (2004) 49--54},
  \href{http://arxiv.org/abs/nucl-th/0305100}{{\ttfamily
  arXiv:nucl-th/0305100}}.

\bibitem{Garcia-Recio2}
D.~Gamermann, C.~Garcia-Recio, J.~Nieves, and L.~L. Salcedo, ``{Odd Parity
  Light Baryon Resonances}'',
  \href{http://dx.doi.org/10.1103/PhysRevD.84.056017}{{\em Phys. Rev. D}
  {\bfseries 84} (2011) 056017},
  \href{http://arxiv.org/abs/1104.2737}{{\ttfamily arXiv:1104.2737 [hep-ph]}}.

\bibitem{Sekihara:2015qqa}
T.~Sekihara, ``{$\Xi (1690)$ as a $\bar{K} \Sigma$ molecular state}'',
  \href{http://dx.doi.org/10.1093/ptep/ptv129}{{\em PTEP} {\bfseries 2015}
  no.~9, (2015) 091D01}, \href{http://arxiv.org/abs/1505.02849}{{\ttfamily
  arXiv:1505.02849 [hep-ph]}}.

\bibitem{Khemchandani:2016ftn}
K.~P. Khemchandani, A.~Mart\'\i{}nez~Torres, A.~Hosaka, H.~Nagahiro, F.~S.
  Navarra, and M.~Nielsen, ``{Why $\Xi(1690)$ and $\Xi(2120)$ are so
  narrow?}'', \href{http://dx.doi.org/10.1103/PhysRevD.97.034005}{{\em Phys.
  Rev. D} {\bfseries 97} no.~3, (2018) 034005},
  \href{http://arxiv.org/abs/1608.07086}{{\ttfamily arXiv:1608.07086
  [nucl-th]}}.

\bibitem{Nishibuchi:2023acl}
T.~Nishibuchi and T.~Hyodo, ``{Analysis of $\Xi(1620)$ resonance and
  $\bar{K}\Lambda$ scattering length with chiral unitary approach}'',
  \href{http://arxiv.org/abs/2305.10753}{{\ttfamily arXiv:2305.10753
  [hep-ph]}}.

\bibitem{Feijoo:2023wua}
A.~Feijoo, V.~Valcarce~Cadenas, and V.~K. Magas, ``{The \ensuremath{\Xi}(1620)
  and \ensuremath{\Xi}(1690) molecular states from S=\ensuremath{-}2
  meson-baryon interaction up to next-to-leading order}'',
  \href{http://dx.doi.org/10.1016/j.physletb.2023.137927}{{\em Phys. Lett. B}
  {\bfseries 841} (2023) 137927},
  \href{http://arxiv.org/abs/2303.01323}{{\ttfamily arXiv:2303.01323
  [hep-ph]}}.

\bibitem{LHCb}
{\bfseries LHCb} Collaboration, R.~Aaij {\em et~al.}, ``{{Evidence of a
  $J/\Psi\Lambda$ structure and observation of excited $\Xi^-$ states in the
  $\Xi^-_b\rightarrow J/\Psi\Lambda K^-$ decay}}'',
  \href{http://dx.doi.org/10.1016/j.scib.2021.02.030}{{\em Sci. Bull.}
  {\bfseries 66} (2021) 1278}.

\bibitem{Belle:2001hyr}
{\bfseries Belle} Collaboration, K.~Abe {\em et~al.}, ``{Observation of Cabibbo
  suppressed and W exchange Lambda+(c) baryon decays}'',
  \href{http://dx.doi.org/10.1016/S0370-2693(01)01373-9}{{\em Phys. Lett. B}
  {\bfseries 524} (2002) 33--43},
  \href{http://arxiv.org/abs/hep-ex/0111032}{{\ttfamily arXiv:hep-ex/0111032}}.

\bibitem{BESIII}
{\bfseries BESIII} Collaboration, M.~Ablikim {\em et~al.}, ``{Study of excited
  $\Xi$ states in
  $\psi(3686)\rightarrow{}K^{-}\Lambda\overline{\Xi}^{+}+c.c.$}'',
  \href{http://arxiv.org/abs/2308.15206}{{\ttfamily arXiv:2308.15206
  [hep-ex]}}.

\bibitem{BELLE}
{\bfseries BELLE} Collaboration, M.~Sumihama {\em et~al.}, ``{{Observation of
  $\Xi(1620)^0$ and $\Xi(1690)^0$ in $\Xi_c^+\rightarrow \Xi^-\pi^+\pi^+$
  decays}}'', \href{http://dx.doi.org/10.1103/PhysRevLett.122.072501}{{\em
  Phys. Rev. Lett.} {\bfseries 122} (2019) 072501}.

\bibitem{Humphrey:1962zz}
W.~E. Humphrey and R.~R. Ross, ``{Low-energy interactions of K$^-$ mesons in
  hydrogen}'',
\href{http://dx.doi.org/10.1103/PhysRev.127.1305}{{\em Phys. Rev.} {\bfseries
  127} (1962) 1305--1323}.

\bibitem{Watson:1963zz}
M.~B. Watson, M.~Ferro-Luzzi, and R.~D. Tripp, ``{Analysis of Y$_0^*$(1520) and
  determination of the $\Sigma$ parity}'',
\href{http://dx.doi.org/10.1103/PhysRev.131.2248}{{\em Phys. Rev.} {\bfseries
  131} (1963) 2248--2281}.

\bibitem{Mast:1975pv}
T.~S. Mast, M.~Alston-Garnjost, R.~O. Bangerter, A.~S. Barbaro-Galtieri, F.~T.
  Solmitz, and R.~D. Tripp, ``{Elastic, Charge Exchange, and Total K- p
  Cross-Sections in the Momentum Range 220-MeV/c to 470-MeV/c}'',
  \href{http://dx.doi.org/10.1103/PhysRevD.14.13}{{\em Phys. Rev. D} {\bfseries
  14} (1976) 13}.

\bibitem{Nowak:1978au}
R.~J. Nowak {\em et~al.}, ``{Charged $\Sigma$ hyperon production by K$^-$ meson
  interactions at rest}'',
\href{http://dx.doi.org/10.1016/0550-3213(78)90179-7}{{\em Nucl. Phys.}
  {\bfseries B139} (1978) 61--71}.

\bibitem{Ciborowski:1982et}
J.~Ciborowski {\em et~al.}, ``{Kaon scattering and charged $\Sigma$ hyperon
  production in K$^-$p interactions below 300~MeV/$c$}'',
\href{http://dx.doi.org/10.1088/0305-4616/8/1/005}{{\em J. Phys.} {\bfseries
  G8} (1982) 13--32}.

\bibitem{Sakitt:1965kh}
M.~Sakitt, T.~B. Day, R.~G. Glasser, N.~Seeman, J.~H. Friedman, W.~E. Humphrey,
  and R.~R. Ross, ``{Low-energy $K^{-}$ meson interactions in Hydrogen}'',
\href{http://dx.doi.org/10.1103/PhysRev.139.B719}{{\em Phys. Rev.} {\bfseries
  139} (1965) B719}.

\bibitem{Bazzi11}
M.~Bazzi {\em et~al.}, ``{{A New Measurement of Kaonic Hydrogen X-rays}}'',
  \href{http://dx.doi.org/10.1016/j.physletb.2011.09.011}{{\em Phys. Lett. B}
  {\bfseries 704} (2011) 113}.

\bibitem{Bazzi12}
M.~Bazzi {\em et~al.}, ``{{Kaonic hydrogen X-ray measurement in SIDDHARTA}}'',
  \href{http://dx.doi.org/10.1016/j.nuclphysa.2011.12.008}{{\em Nucl. Phys. A}
  {\bfseries 881} (2012) 88}.

\bibitem{Lambda1405_1}
V.~K. Magas, E.~Oset, and A.~Ramos, ``{Evidence for the two pole structure of
  the $\Lambda$(1405) resonance}'',
  \href{http://dx.doi.org/10.1103/PhysRevLett.95.052301}{{\em Phys. Rev. Lett.}
  {\bfseries 95} (2005) 052301},
  \href{http://arxiv.org/abs/hep-ph/0503043}{{\ttfamily arXiv:hep-ph/0503043}}.

\bibitem{Lambda1405_2}
J.~Siebenson and L.~Fabbietti, ``{Investigation of the
  \ensuremath{\Lambda}(1405) line shape observed in pp collisions}'',
  \href{http://dx.doi.org/10.1103/PhysRevC.88.055201}{{\em Phys. Rev. C}
  {\bfseries 88} (2013) 055201},
  \href{http://arxiv.org/abs/1306.5183}{{\ttfamily arXiv:1306.5183 [nucl-ex]}}.

\bibitem{Lambda1405_3}
R.~J. Hemingway, ``{Production of $\Lambda(1405)$ in K$^-$p reactions at 4.2
  {GeV}/$c$}'', \href{http://dx.doi.org/10.1016/0550-3213(85)90556-5}{{\em
  Nucl. Phys.} {\bfseries B253} (1985) 742--752}.

\bibitem{Lambda1405_4}
I.~Zychor {\em et~al.}, ``{Shape of the $\Lambda$(1405) hyperon measured
  through its $\Sigma^0\pi^0$ decay}'',
  \href{http://dx.doi.org/10.1016/j.physletb.2008.01.002}{{\em Phys. Lett. B}
  {\bfseries 660} (2008) 167--171},
  \href{http://arxiv.org/abs/0705.1039}{{\ttfamily arXiv:0705.1039 [nucl-ex]}}.

\bibitem{Lambda1405_CLAS}
{\bfseries CLAS} Collaboration, K.~Moriya and R.~Schumacher, ``{Properties of
  the $\Lambda$(1405) measured at CLAS}'',
  \href{http://dx.doi.org/10.1016/j.nuclphysa.2010.01.210}{{\em Nucl. Phys.}
  {\bfseries A835} (2010) 325--328},
  \href{http://arxiv.org/abs/0911.2705}{{\ttfamily arXiv:0911.2705 [nucl-ex]}}.

\bibitem{LKALICEpp}
{\bfseries ALICE} Collaboration, S.~Acharya {\em et~al.}, ``{Accessing the
  strong interaction between \ensuremath{\Lambda} baryons and charged kaons
  with the femtoscopy technique at the LHC}'',
  \href{http://dx.doi.org/10.1016/j.physletb.2023.138145}{{\em Phys. Lett. B}
  {\bfseries 845} (2023) 138145},
  \href{http://arxiv.org/abs/2305.19093}{{\ttfamily arXiv:2305.19093
  [nucl-ex]}}.

\bibitem{ALICE:Run1}
{\bfseries ALICE} Collaboration, S.~Acharya {\em et~al.}, ``{p-p, p-$\Lambda$
  and $\Lambda$-$\Lambda$ correlations studied via femtoscopy in pp reactions
  at $\sqrt{s}$ = 7 TeV}'',
  \href{http://dx.doi.org/10.1103/PhysRevC.99.024001}{{\em Phys. Rev.}
  {\bfseries C99} (2019) 024001},
  \href{http://arxiv.org/abs/1805.12455}{{\ttfamily arXiv:1805.12455
  [nucl-ex]}}.

\bibitem{ALICE:pXi}
{\bfseries ALICE} Collaboration, S.~Acharya {\em et~al.}, ``{First Observation
  of an Attractive Interaction between a Proton and a Cascade Baryon}'',
  \href{http://dx.doi.org/10.1103/PhysRevLett.123.112002}{{\em Phys. Rev.
  Lett.} {\bfseries 123} (2019) 112002},
  \href{http://arxiv.org/abs/1904.12198}{{\ttfamily arXiv:1904.12198
  [nucl-ex]}}.

\bibitem{ALICE:pSig0}
{\bfseries ALICE} Collaboration, S.~Acharya {\em et~al.}, ``{Investigation of
  the p-$\Sigma^{0}$ interaction via femtoscopy in pp collisions}'',
  \href{http://dx.doi.org/10.1016/j.physletb.2020.135419}{{\em Phys. Lett. \bf
  B} {\bfseries 805} (2020) 135419},
  \href{http://arxiv.org/abs/1910.14407}{{\ttfamily arXiv:1910.14407
  [nucl-ex]}}.

\bibitem{ALICE:LL}
{\bfseries ALICE} Collaboration, S.~Acharya {\em et~al.}, ``{Study of the
  $\Lambda$-$\Lambda$ interaction with femtoscopy correlations in pp and p--Pb
  collisions at the LHC}'',
  \href{http://dx.doi.org/10.1016/j.physletb.2019.134822}{{\em Phys. Lett. \bf
  B} {\bfseries 797} (2019) 134822},
  \href{http://arxiv.org/abs/1905.07209}{{\ttfamily arXiv:1905.07209
  [nucl-ex]}}.

\bibitem{ALICE:pOmega}
{\bfseries ALICE} Collaboration, S.~Acharya {\em et~al.}, ``{Unveiling the
  strong interaction among hadrons at the LHC}'',
  \href{http://dx.doi.org/10.1038/s41586-020-3001-6}{{\em Nature} {\bfseries
  588} no.~7837, (2020) 232--238},
  \href{http://arxiv.org/abs/2005.11495}{{\ttfamily arXiv:2005.11495
  [nucl-ex]}}.

\bibitem{ALICE:pKpp}
{\bfseries ALICE} Collaboration, S.~Acharya {\em et~al.}, ``{{Scattering
  Studies with Low-Energy Kaon-Proton Femtoscopy in Proton-Proton Collisions at
  the LHC}}'', \href{http://dx.doi.org/10.1103/PhysRevLett.124.092301}{{\em
  Phys. Rev. Lett.} {\bfseries 124} (2020) 092301}.

\bibitem{ALICE:pKCoupled}
{\bfseries ALICE} Collaboration, S.~Acharya {\em et~al.}, ``{Constraining the
  ${\rm\overline{K}N}$ coupled channel dynamics using femtoscopic correlations
  at the LHC}'', \href{http://dx.doi.org/10.1140/epjc/s10052-023-11476-0}{{\em
  Eur.Phys.J.C} {\bfseries 83} (5, 2022) },
  \href{http://arxiv.org/abs/2205.15176}{{\ttfamily arXiv:2205.15176
  [nucl-ex]}}.

\bibitem{ALICE:pLCoupled}
{\bfseries ALICE} Collaboration, S.~Acharya {\em et~al.}, ``{Exploring the
  N\ensuremath{\Lambda}\textendash{}N\ensuremath{\Sigma} coupled system with
  high precision correlation techniques at the LHC}'',
  \href{http://dx.doi.org/10.1016/j.physletb.2022.137272}{{\em Phys. Lett. B}
  {\bfseries 833} (2022) 137272},
  \href{http://arxiv.org/abs/2104.04427}{{\ttfamily arXiv:2104.04427
  [nucl-ex]}}.

\bibitem{ALICE:BBar}
{\bfseries ALICE} Collaboration, S.~Acharya {\em et~al.}, ``{Investigating the
  role of strangeness in baryon\textendash{}antibaryon annihilation at the
  LHC}'', \href{http://dx.doi.org/10.1016/j.physletb.2022.137060}{{\em Phys.
  Lett. B} {\bfseries 829} (2022) 137060},
  \href{http://arxiv.org/abs/2105.05190}{{\ttfamily arXiv:2105.05190
  [nucl-ex]}}.

\bibitem{ALICE:LXi}
{\bfseries ALICE} Collaboration, ``{First measurement of the $\Lambda$-$\Xi$
  interaction in proton-proton collisions at the LHC}'',
  \href{http://arxiv.org/abs/2204.10258}{{\ttfamily arXiv:2204.10258
  [nucl-ex]}}.

\bibitem{ALICERun3Run4}
{\bfseries ALICE} Collaboration, ``{Future high-energy pp programme with
  ALICE}'',.

\bibitem{ALICE3}
{\bfseries ALICE} Collaboration, ``{Letter of intent for ALICE 3: A
  next-generation heavy-ion experiment at the LHC}'',
  \href{http://arxiv.org/abs/2211.02491}{{\ttfamily arXiv:2211.02491
  [physics.ins-det]}}.

\bibitem{LisaPrattReview}
M.~A. Lisa, S.~Pratt, R.~Soltz, and U.~Wiedemann, ``{Femtoscopy in relativistic
  heavy ion collisions}'',
  \href{http://dx.doi.org/10.1146/annurev.nucl.55.090704.151533}{{\em Ann. Rev.
  Nucl. Part. Sci.} {\bfseries 55} (2005) 357--402},
  \href{http://arxiv.org/abs/nucl-ex/0505014}{{\ttfamily
  arXiv:nucl-ex/0505014}}.

\bibitem{Haidenbauer:CC}
J.~Haidenbauer, ``{Coupled-channel effects in hadron-hadron correlation
  functions}'', \href{http://dx.doi.org/10.1016/j.nuclphysa.2018.10.090}{{\em
  Nucl. Phys.} {\bfseries A981} (2019) 1--16},
  \href{http://arxiv.org/abs/1808.05049}{{\ttfamily arXiv:1808.05049
  [hep-ph]}}.

\bibitem{Kamiya:CCKp}
Y.~Kamiya, T.~Hyodo, K.~Morita, A.~Ohnishi, and W.~Weise, ``{$K^-p$ Correlation
  Function from High-Energy Nuclear Collisions and Chiral SU(3) Dynamics}'',
  \href{http://dx.doi.org/10.1103/PhysRevLett.124.132501}{{\em Phys. Rev.
  Lett.} {\bfseries 124} no.~13, (2020) 132501},
  \href{http://arxiv.org/abs/1911.01041}{{\ttfamily arXiv:1911.01041
  [nucl-th]}}.

\bibitem{ALICE:Source}
{\bfseries ALICE} Collaboration, S.~Acharya {\em et~al.}, ``{Search for a
  common baryon source in high-multiplicity pp collisions at the LHC}'',
  \href{http://dx.doi.org/10.1016/j.physletb.2020.135849}{{\em Phys. Lett. B}
  {\bfseries 811} (2020) 135849},
  \href{http://arxiv.org/abs/2004.08018}{{\ttfamily arXiv:2004.08018
  [nucl-ex]}}.

\bibitem{CECA}
D.~Mihaylov and J.~Gonz\'alez~Gonz\'alez, ``{Novel model for particle emission
  in small collision systems}'',
  \href{http://dx.doi.org/10.1140/epjc/s10052-023-11774-7}{{\em Eur. Phys. J.
  C} {\bfseries 83} no.~7, (2023) 590},
  \href{http://arxiv.org/abs/2305.08441}{{\ttfamily arXiv:2305.08441
  [hep-ph]}}.

\bibitem{Feijoo:2021zau}
A.~Feijoo, D.~Gazda, V.~Magas, and A.~Ramos, ``{The K\textasciimacron{}N
  Interaction in Higher Partial Waves}'',
  \href{http://dx.doi.org/10.3390/sym13081434}{{\em Symmetry} {\bfseries 13}
  no.~8, (2021) 1434}, \href{http://arxiv.org/abs/2107.10560}{{\ttfamily
  arXiv:2107.10560 [hep-ph]}}.

\bibitem{Ramos:2016odk}
A.~Ramos, A.~Feijoo, and V.~K. Magas, ``{The chiral S = \ensuremath{-}1
  meson\textendash{}baryon interaction with new constraints on the NLO
  contributions}'',
  \href{http://dx.doi.org/10.1016/j.nuclphysa.2016.05.006}{{\em Nucl. Phys. A}
  {\bfseries 954} (2016) 58--74},
  \href{http://arxiv.org/abs/1605.03767}{{\ttfamily arXiv:1605.03767
  [nucl-th]}}.

\bibitem{femtoreview}
L.~Fabbietti, V.~M. Sarti, and O.~V. Doce, ``{Study of the strong interaction
  among hadrons with correlations at the LHC}'',
  \href{http://dx.doi.org/10.1146/annurev-nucl-102419-034438}{{\em Ann. Rev.
  Nucl. Part. Sci.} {\bfseries 71} (2021) 377--402},
  \href{http://arxiv.org/abs/2012.09806}{{\ttfamily arXiv:2012.09806
  [nucl-ex]}}.

\bibitem{ALICE:pD}
{\bfseries ALICE} Collaboration, S.~Acharya {\em et~al.}, ``{First study of the
  two-body scattering involving charm hadrons}'',
  \href{http://dx.doi.org/10.1103/PhysRevD.106.052010}{{\em Phys. Rev. D}
  {\bfseries 106} no.~5, (2022) 052010},
  \href{http://arxiv.org/abs/2201.05352}{{\ttfamily arXiv:2201.05352
  [nucl-ex]}}.

\bibitem{TFMultiplicity}
V.~Vovchenko, B.~D\"onigus, and H.~Stoecker, ``{Canonical statistical model
  analysis of pp , p--Pb, and Pb--Pb collisions at energies available at the
  CERN Large Hadron Collider}'',
  \href{http://dx.doi.org/10.1103/PhysRevC.100.054906}{{\em Phys. Rev. C}
  {\bfseries 100} no.~5, (2019) 054906},
  \href{http://arxiv.org/abs/1906.03145}{{\ttfamily arXiv:1906.03145
  [hep-ph]}}.

\bibitem{hepdata.143518}
{ALICE Collaboration}, ``{Accessing the strong interaction between $\Lambda$
  baryons and charged kaons with the femtoscopy technique at the LHC}.''
  {HEPData (collection)}, 2023.
\newblock \url{https://doi.org/10.17182/hepdata.143518}.

\bibitem{PDG}
{\bfseries Particle Data Group} Collaboration, R.~L. Workman {\em et~al.},
  ``{Review of Particle Physics}'',
  \href{http://dx.doi.org/10.1093/ptep/ptac097}{{\em PTEP} {\bfseries 2022}
  (2022) 083C01}.

\bibitem{Oller:2000fj}
J.~A. Oller and U.~G. Meissner, ``{Chiral dynamics in the presence of bound
  states: Kaon nucleon interactions revisited}'',
  \href{http://dx.doi.org/10.1016/S0370-2693(01)00078-8}{{\em Phys. Lett. B}
  {\bfseries 500} (2001) 263--272},
  \href{http://arxiv.org/abs/hep-ph/0011146}{{\ttfamily arXiv:hep-ph/0011146}}.

\bibitem{Borasoy:2005ie}
B.~Borasoy, R.~Nissler, and W.~Weise, ``{Chiral dynamics of kaon-nucleon
  interactions, revisited}'',
  \href{http://dx.doi.org/10.1140/epja/i2005-10079-1}{{\em Eur. Phys. J. A}
  {\bfseries 25} (2005) 79--96},
  \href{http://arxiv.org/abs/hep-ph/0505239}{{\ttfamily arXiv:hep-ph/0505239}}.

\bibitem{Guo:2012vv}
Z.-H. Guo and J.~A. Oller, ``{Meson-baryon reactions with strangeness -1 within
  a chiral framework}'',
  \href{http://dx.doi.org/10.1103/PhysRevC.87.035202}{{\em Phys. Rev. C}
  {\bfseries 87} no.~3, (2013) 035202},
  \href{http://arxiv.org/abs/1210.3485}{{\ttfamily arXiv:1210.3485 [hep-ph]}}.

\bibitem{Feijoo:2018den}
A.~Feijoo, V.~Magas, and A.~Ramos, ``{$S$=\ensuremath{-}1 meson-baryon
  interaction and the role of isospin filtering processes}'',
  \href{http://dx.doi.org/10.1103/PhysRevC.99.035211}{{\em Phys. Rev. C}
  {\bfseries 99} no.~3, (2019) 035211},
  \href{http://arxiv.org/abs/1810.07600}{{\ttfamily arXiv:1810.07600
  [hep-ph]}}.

\bibitem{PresTeukVettFlan92}
W.~H. Press, S.~A. Teukolsky, W.~T. Vetterling, and B.~P. Flannery, {\em
  Numerical Recipes in C}.
\newblock Cambridge University Press, Cambridge, USA, second~ed., 1992.

\bibitem{Efron:1986hys}
B.~Efron and R.~Tibshirani, ``{An introduction to the bootstrap}'', {\em
  Statist. Sci.} {\bfseries 57} no.~1, (1986) 54--75.

\bibitem{Albaladejo:2016hae}
M.~Albaladejo, D.~Jido, J.~Nieves, and E.~Oset, ``{$D^*_{s0}(2317)$ and
  $\textit{DK}$ scattering in B decays from BaBar and LHCb data}'',
  \href{http://dx.doi.org/10.1140/epjc/s10052-016-4144-3}{{\em Eur. Phys. J. C}
  {\bfseries 76} no.~6, (2016) 300},
  \href{http://arxiv.org/abs/1604.01193}{{\ttfamily arXiv:1604.01193
  [hep-ph]}}.

\bibitem{Braaten:2004rn}
E.~Braaten and H.~W. Hammer, ``{Universality in few-body systems with large
  scattering length}'',
  \href{http://dx.doi.org/10.1016/j.physrep.2006.03.001}{{\em Phys. Rept.}
  {\bfseries 428} (2006) 259--390},
  \href{http://arxiv.org/abs/cond-mat/0410417}{{\ttfamily
  arXiv:cond-mat/0410417}}.

\bibitem{Naidon:2016dpf}
P.~Naidon and S.~Endo, ``{Efimov Physics: a review}'',
  \href{http://dx.doi.org/10.1088/1361-6633/aa50e8}{{\em Rept. Prog. Phys.}
  {\bfseries 80} no.~5, (2017) 056001},
  \href{http://arxiv.org/abs/1610.09805}{{\ttfamily arXiv:1610.09805
  [quant-ph]}}.

\bibitem{Hyodo:2013iga}
T.~Hyodo, ``{Structure of Near-Threshold s-Wave Resonances}'',
  \href{http://dx.doi.org/10.1103/PhysRevLett.111.132002}{{\em Phys. Rev.
  Lett.} {\bfseries 111} (2013) 132002},
  \href{http://arxiv.org/abs/1305.1999}{{\ttfamily arXiv:1305.1999 [hep-ph]}}.

\bibitem{Giacosa:2021mbz}
F.~Giacosa, A.~Okopi\'nska, and V.~Shastry, ``{A simple alternative to the
  relativistic Breit\textendash{}Wigner distribution}'',
  \href{http://dx.doi.org/10.1140/epja/s10050-021-00641-2}{{\em Eur. Phys. J.
  A} {\bfseries 57} no.~12, (2021) 336},
  \href{http://arxiv.org/abs/2106.03749}{{\ttfamily arXiv:2106.03749
  [hep-ph]}}.

\end{thebibliography}\endgroup

\end{document}